\documentstyle[ijmpal,epsfig,epsf,twoside]{article}
\newcommand{\DO}{D\raise1pt\hbox{$\not$}O}
\begin{document}
\runninghead{Study of direct-photon and pion production} 
{Study of direct-photon and pion production}

\normalsize\textlineskip
\thispagestyle{empty}
\setcounter{page}{1}

\copyrightheading{}                     

\vspace*{0.88truein}

\fpage{1}
\centerline{\bf STUDY OF DIRECT-PHOTON AND PION PRODUCTION}
\vspace*{0.37truein}
\centerline{\footnotesize MAREK ZIELI\'NSKI}
\vspace*{0.015truein}
\centerline{\footnotesize\it University of Rochester}
\baselineskip=10pt
\centerline{\footnotesize\it Rochester, NY 14627, USA}

\vspace*{0.21truein}

\abstracts{
We present comparisons of high-$p_T$ photon  and pion production in
hadronic interactions with expectations from next-to-leading order
perturbative QCD (NLO pQCD). We also comment on
phenomenological models of $k_T$ smearing (which approximate
effects of additional soft-gluon emission) 
and on the status of resummation calculations.  
}{}{}
\textlineskip                  
\vspace*{12pt}                 

\noindent
Direct-photon production in hadronic collisions at high transverse
momenta ($p_T$) has long been viewed as an ideal testing ground for
the formalism of pQCD.  A reliable theoretical description of the
direct-photon process is of special importance because of its
sensitivity to the gluon distribution in a proton through the
quark--gluon scattering subprocess ($gq\rightarrow\gamma q$).  The
gluon distribution, $G(x)$, is relatively well constrained for $x<0.25$,
but much less so at larger~$x$.
In principle, fixed-target direct-photon production can constrain $G(x)$
at large~$x$, and such data have therefore been incorporated in
fits to global parton distribution functions (PDF).

\begin{figure}
\vspace{-.3in}
\centerline{
\hspace{0.3in}
\epsfxsize=2.2truein
\epsffile{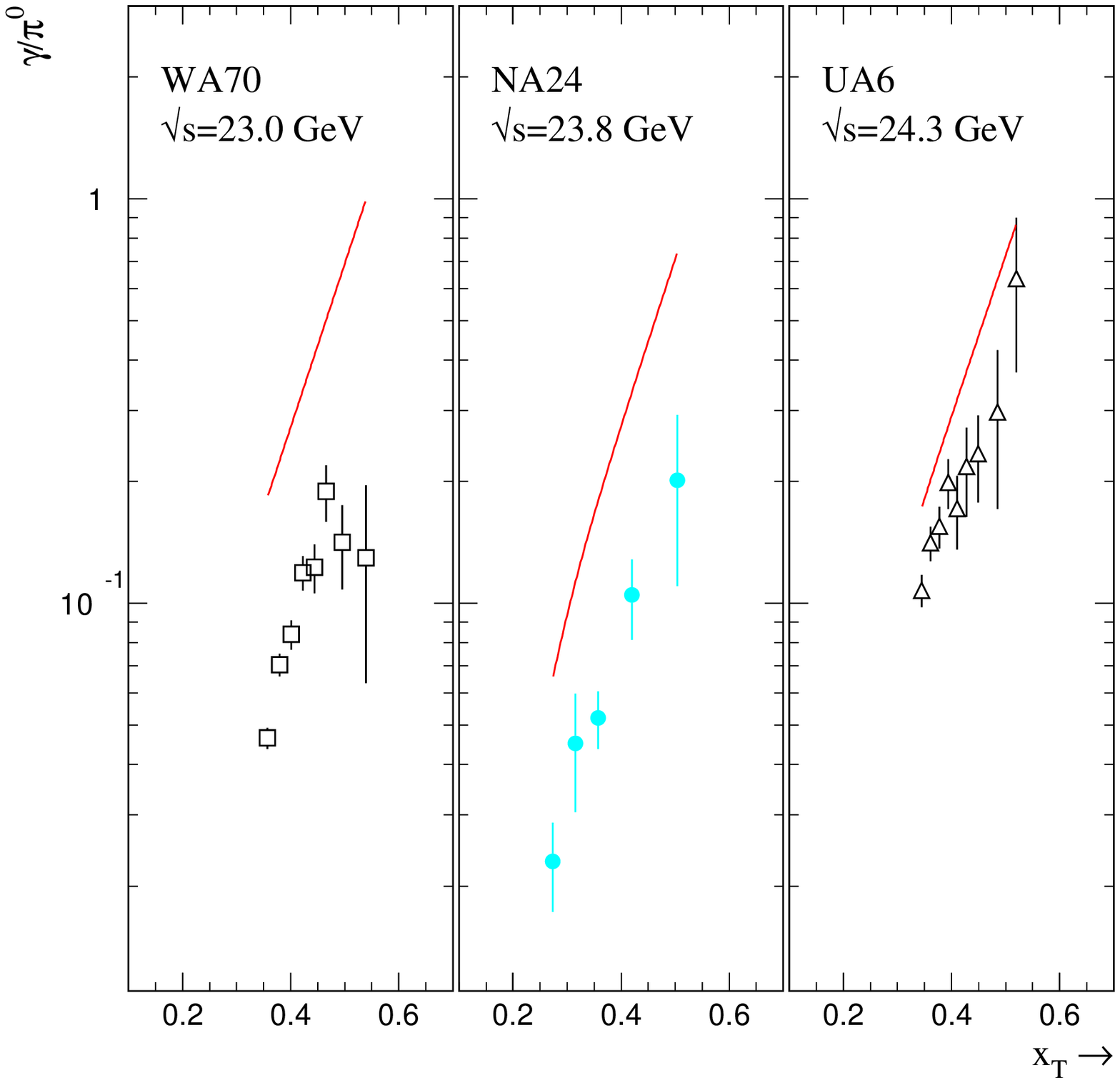} 
\hspace{0.3in}
\epsfxsize=2.2truein
\epsffile{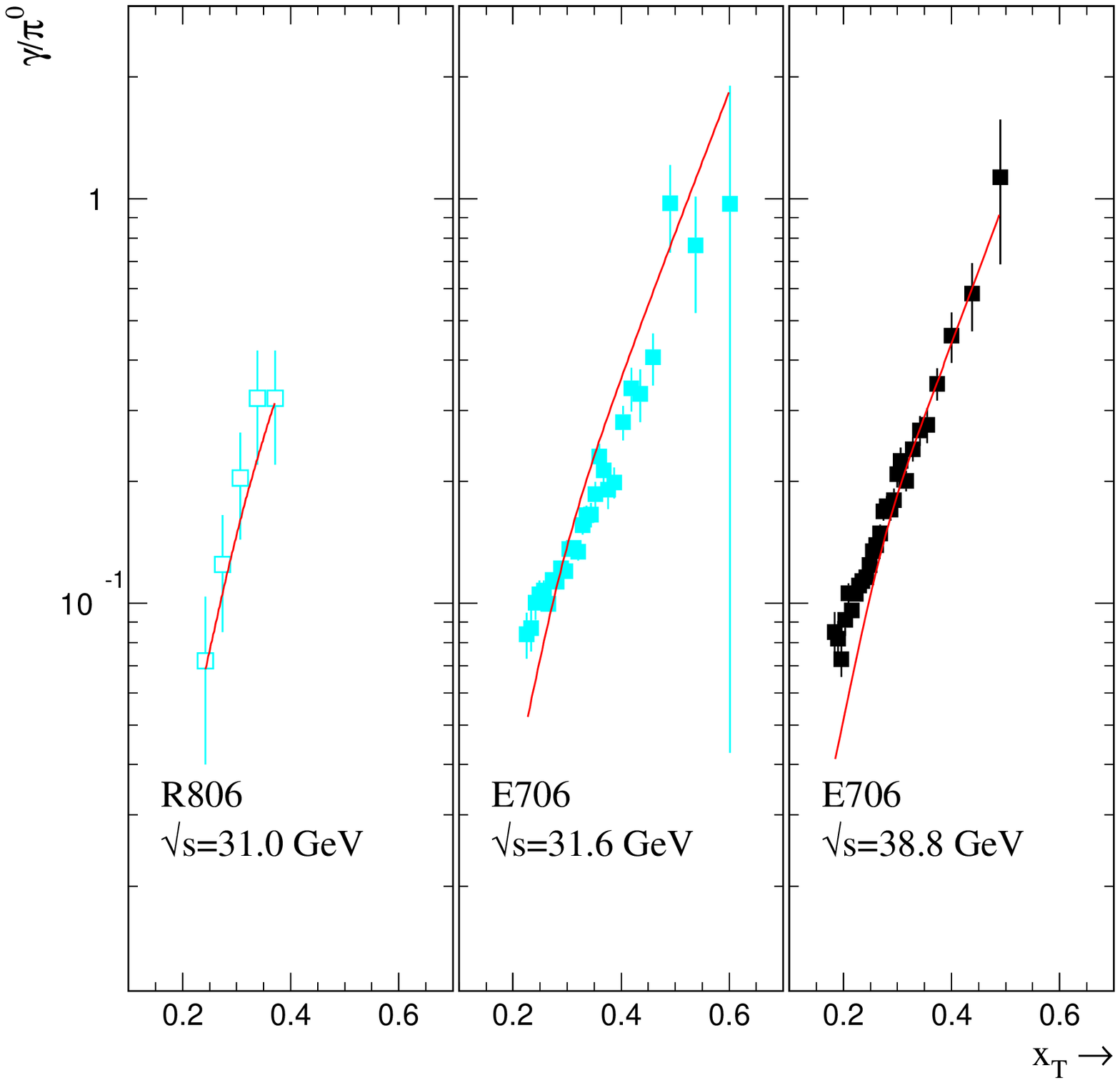} 
}
\vspace{-.5in}
\caption{
Left:
Comparison of $\gamma/\pi^0$ rates 
as a function of $x_T$ for WA70, NA24, and UA6 at
$\sqrt{s}\approx23-24$~GeV. 
Overlayed are the results from NLO pQCD.
Right:
Same for R806 at $\sqrt{s}=31$~GeV and E706 at $\sqrt{s}=31.6$ and $38.8$~GeV.
\label{fig:photonTOpi0_1}}
\end{figure}

\begin{figure}
\centering\leavevmode
\vglue1truept\vspace{-0.25in}
\hglue1truept\hspace{-2.3in}
\epsfxsize=2.3truein
\epsffile{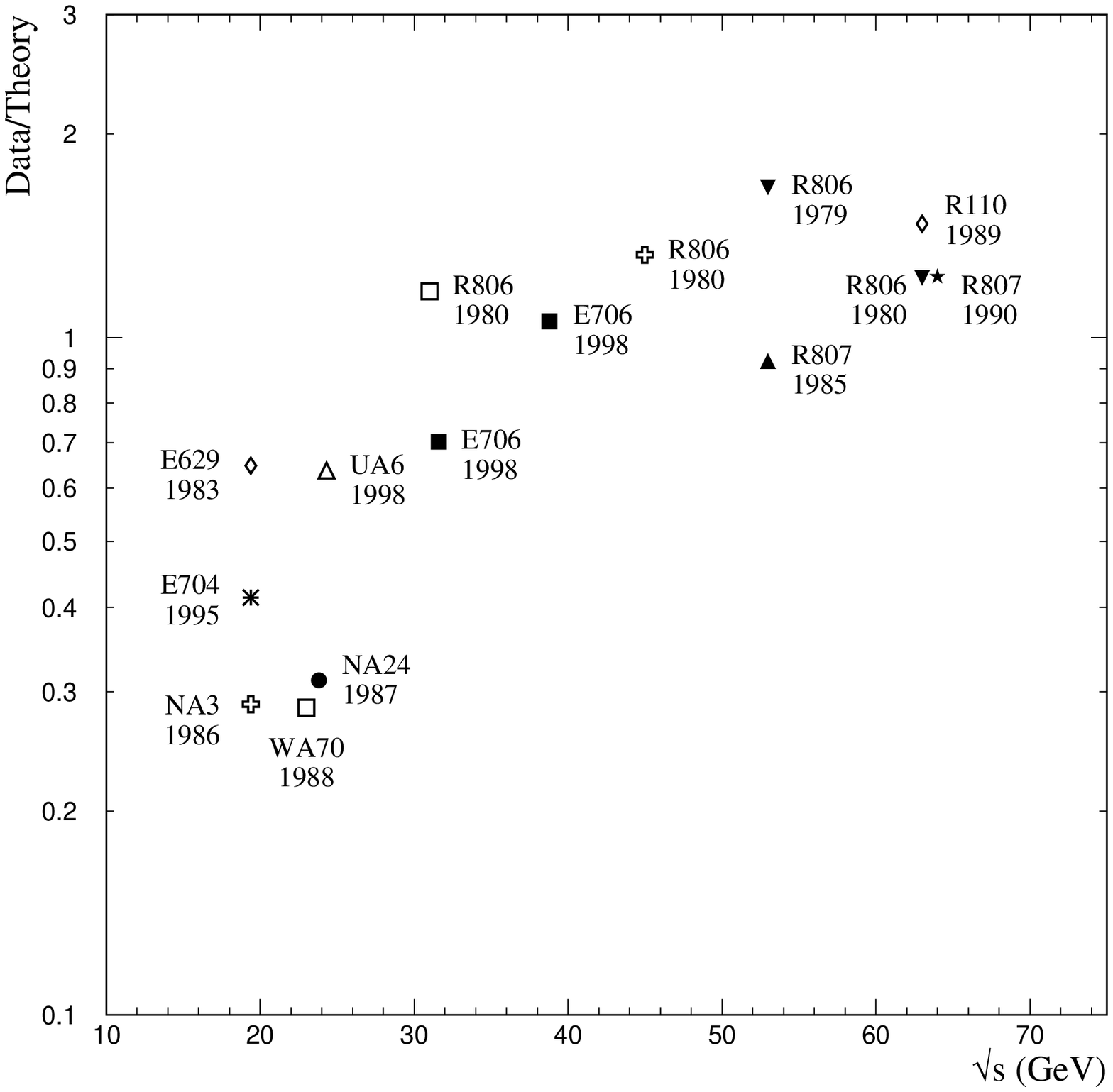}
\vskip -7.5cm
\hglue1truept\hskip 2.6in
\epsfig{figure=e706crs.eps,height=2.3truein,angle=90}
\vskip -.5cm
\caption{Left:
The ratio of data for the $\gamma/\pi^0$ measurements to NLO theory,  
as a function of $\sqrt{s}$.
Right:
Comparison between the E706 direct-photon data\protect\cite{E706-kt} at 
$\sqrt{s}=31.6$~GeV 
and recent pQCD calculations. The dotted
line represents the full NLO calculation,\protect\cite{gordon} while
the dashed and solid lines, respectively, incorporate purely threshold
resummation\protect\cite{nason} and joint threshold and recoil
resummation.\protect\cite{sterman}
\label{fig:photonTOpi0_2}}
\end{figure}

However, both the completeness of the NLO description of
the direct-$\gamma$ process, as well as the consistency of results from
different experiments, have been 
questioned.\cite{mrst,huston-discrepancy,E706-kt,ktprd,aurenche-dp,kimber}
The inclusive production of hadrons provides a further means for
testing predictions of the NLO pQCD formalism.  Deviations have, in fact,
been observed between measured inclusive direct-$\gamma$ and pion production
cross sections and NLO pQCD calculations, and it has been suggested that 
part of this discrepancy can be ascribed to higher-order effects of
initial-state soft-gluon radiation.\cite{huston-discrepancy,E706-kt,ktprd} 
However, it seems unlikely that theoretical developments alone will be able 
to accommodate the observed level of scatter in the ratio of data to theory
for $\gamma$ and $\pi^0$ yields. These observations motivated us
to consider measurements of the $\gamma/\pi^0$ ratio 
over a wide range of $\sqrt{s}$.\cite{osaka}  
Both experimental and theoretical uncertainties tend
to cancel in such a ratio, and the ratio should also be less
sensitive to incomplete treatment of gluon radiation.
A sample of the ratio of direct-$\gamma$ to $\pi^0$ cross sections for 
both data and NLO pQCD is given in 
Fig.~\ref{fig:photonTOpi0_1}
for incident protons, as a function of $x_T=2p_T/\sqrt{s}$, for   
$\sqrt{s}\approx 23-24$~GeV (left) and  $\sqrt{s}=31-39$~GeV (right).
For all measurements in the left panel, 
theory is high compared to data; similar levels of agreement are observed
at $\sqrt{s}=19.4$~GeV.\cite{osaka} 
At larger $\sqrt{s}$, the NLO value for the ratio agrees better with 
experiment, as seen in Fig.~\ref{fig:photonTOpi0_1} (right) for 
$\sqrt{s}=31-39$~GeV.
At even higher $\sqrt{s}$, theory lies slightly below the data.\cite{osaka} 
(The NLO calculations use a single scale of $\mu=p_T/2$, 
CTEQ4M PDFs,\cite{cteq4} and BKK fragmentation functions\cite{BKK} for 
pions.) 
A compilation of these results, shown for simplicity without their
uncertainties, is presented in Fig.~\ref{fig:photonTOpi0_2} (left).  
Here, the ratios of data to theory for the ratio of $\gamma$ to $\pi^0$
yields (e.g., in Fig.~\ref{fig:photonTOpi0_1}) were fitted as constant 
values at high $p_T$, and the results plotted as a function of $\sqrt{s}$. 
The figure suggests an energy dependence in the ratio of data to theory 
for the $\gamma/\pi^0$ production ratio.  There are also substantial
differences between the experiments at low $\sqrt{s}$ (where the
observed $\gamma/\pi^0$ is smallest), which makes it difficult to
quantify this trend.  Recognizing the presence of these differences is
especially important because thus far only the low energy photon experiments
have been used in PDF fits to extract $G(x)$.

Recently, an intuitive, and often successful, phenomenological approach 
has been used to describe soft gluon radiation in high-$p_T$ inclusive 
production,\cite{ktprd} parametrized in 
terms of an effective $k_T$ that provided  additional 
transverse impulse to the outgoing partons. At fixed-target energies,  
the resulting corrections for direct-$\gamma$ and $\pi^0$ production 
can be large over the full range of $p_T$.\cite{osaka} The corrections depend 
on the values used for $\langle k_T\rangle$, with changes of 200 MeV/$c$ 
making substantial difference. It is therefore difficult to obtain the 
precision needed for extracting global parton distributions. In addition, 
there are different ways to implement such models,\cite{mrst,osaka} 
which can produce quantitative differences in the $k_T$-correction factors. 

Resummed pQCD calculations for single direct-$\gamma$ production are 
currently under development.\cite{nason,kidonakisowens,sterman}
Two recent threshold-resummed pQCD calculations 
for direct photons\cite{nason,kidonakisowens} exhibit far less dependence 
on QCD scales than NLO theory, and provide an enhancement at high~$p_T$.
A method for simultaneous treatment of recoil and threshold
corrections in inclusive single-$\gamma$ cross sections is also being
developed.\cite{sterman} This approach accounts explicitly for the 
recoil from soft radiation in the hard scattering, and 
conserves both energy and transverse momentum for the resummed radiation.  
The possibility of substantial enhancements from higher-order perturbative
and power-law nonperturbative corrections relative to NLO are
indicated at both moderate and high $p_T$ for fixed-target energies
(Fig.~\ref{fig:photonTOpi0_2} (right)), similar to the enhancements 
obtained with simple $k_T$-smearing.\cite{ktprd} 
These recent developments in theory of direct-$\gamma$ processes provide 
cause for optimism that the long-standing difficulties in developing an 
adequate description of direct-$\gamma$ production can eventually be resolved, 
making possible a reliable extraction of $G(x)$ from such 
data.\cite{sterman-pdf}

This work was done in collaboration with L. Apanasevich, 
M.~Begel, C.~Brom\-berg, T.~Ferbel,  G.~Ginther,
J.~Huston, S.~Kuhlmann, P.~Slattery, and V.~Zutshi.
We also wish to thank  P.~Aurenche,  S.~Catani, M.~Fontannaz,
G.~Sterman, and W.~Vogelsang for helpful discussions.

\renewcommand{\baselinestretch}{1.}

\end{document}